\documentclass[aps,twocolumn,prl,superscriptaddress,amsmath,showpacs,tightenlines]{revtex4}
\usepackage{graphicx}
\usepackage{epsfig}

\begin{document}

\title{Producing cluster states in charge qubits and flux qubits}

\author{Tetsufumi Tanamoto}
\affiliation{Corporate R \& D center, Toshiba Corporation,
Saiwai-ku, Kawasaki 212-8582, Japan}

\author{Yu-xi Liu}
\affiliation{Frontier Research System, The Institute of Physical
and Chemical Research (RIKEN), Wako-shi, Saitama 351-0198,
Japan}

\author{Shinobu Fujita}
\affiliation{Corporate R \& D center, Toshiba Corporation,
Saiwai-ku, Kawasaki 212-8582, Japan}

\author{Xuedong Hu}
\affiliation{Department of Physics, University at Buffalo, SUNY,
Buffalo, New York 14260-1500,USA}

\author{Franco Nori}
\affiliation{Frontier Research System, The Institute of Physical
and Chemical Research (RIKEN), Wako-shi, Saitama 351-0198,
Japan}
\affiliation{Physics Department, MCTP, The University of
Michigan, Ann Arbor, Michigan 48109-1040, USA}

\date{\today}

\begin{abstract}
We propose a method to efficiently generate cluster states 
in charge qubits, both semiconducting and superconducting, as 
well as flux qubits.  
We show that highly-entangled cluster states can be 
realized by a `one-touch' entanglement operation by tuning gate bias voltages 
for charge qubits. 
We also investigate the robustness
of these cluster states for non-uniform qubits, which are unavoidable 
in solid-state systems. We find that
quantum computation based on cluster states is a promising
approach for solid-state qubits.
\end{abstract}
\maketitle

One-way quantum computing \cite{Briegel2}, which is based on a series of
one-qubit measurements starting from cluster states of a
qubit array, is an intriguing alternative to the widely 
studied approach using unitary quantum gates. 
Here, the power of quantum mechanics, 
such as quantum parallelism and entanglement, is already stored in the 
initial cluster state. 
Cluster states are fixed highly-entangled states that 
involve all qubits and act as a universal resource for quantum computing.

Because of their unique importance, cluster states have been studied 
in a variety of physical systems.  They have been extensively explored 
in optical quantum computers both theoretically \cite{Nielsen} and 
experimentally \cite{Walther}.  By 
incorporating cluster states, optical quantum computing can achieve 
substantially simpler operations \cite{Nielsen} compared to the original linear optics 
quantum computing proposal \cite{Knill}.  Cluster states have also been 
studied in solid state qubits.  In particular, 
processes of generating cluster states for single and encoded spin qubits 
have been proposed \cite{Loss,Levy} using the Heisenberg exchange 
interaction.  

The existing methods of generating cluster states all require multiple steps
because of the types of interaction involved
(in the case of photonic qubits, a large number of optical elements is
also required).
Here we describe theoretically an efficient method to create scalable 
cluster states in charge 
qubits \cite{Gorman,Hayashi,tana0,Koiller,Nakamura,Makhlin,Yamamoto} 
and flux qubits \cite{Majer,Liu,Rigetti}, 
using existing Ising-like interactions.  
Our key result is that {\it cluster states in charge qubits can be created 
by applying a {\it single} gate bias pulse}, right after preparing an initial 
product state 
$|\Psi_0\rangle \equiv |\Psi(t=0)\rangle = \Pi_i |+\rangle_i\,$, where 
$|\pm \rangle_i=(|0\rangle_i\!\pm\!|1\rangle_i)/\sqrt{2}$.  
We also calculate 
the time-dependent fidelity of the cluster states in charge qubits 
using a quantum dot (QD) system with 
decoherence produced by the measurement back-action, and  
explore the effects of non-uniformity among qubits, which is a realistic 
characteristics for all solid-state qubits.

{\it Cluster states in charge qubits.}--- The Hamiltonian for an
array of charge qubits with nearest-neighbor interactions is described by
\begin{equation}
H_{\rm cq}=\sum_i (\Omega_i \, \sigma_{ix}
+\epsilon_{i}\,\sigma_{iz})
+\sum_{i<j}J_{ij}\,\sigma_{iz}\,\sigma_{jz}
\label{Hcq}
\end{equation}
with Pauli matrices $\sigma_{ix}$ and $\sigma_{iz}$
for the $i$-th qubit.  $\Omega_i$ is either the inter-QD tunnel
coupling for coupled QD systems~\cite{Gorman,Hayashi,tana0}, 
or half the Josephson energy 
for superconducting charge qubits \cite{Nakamura,Makhlin,Yamamoto}, 
respectively.  For either semiconducting or superconducting charge qubits, 
$\epsilon_i$ is the charging energy,
and corresponds to the energy difference 
between $|0\rangle$ and $|1\rangle$ for each qubit.
The coupling constants $J_{ij}$ are derived from the capacitance 
couplings.  In one-way quantum computing \cite{Briegel2},
the $\sigma_{ix}$ term needs to be switched off during the
creation of the cluster state ($H_{\rm cs}$), and then switched on
when measurements are carried out.  From this perspective,
charge qubits with tunable $\Omega_i$ \cite{You} are
desirable.  
However, tunability can produce decoherence and cross-talk
between qubits themselves and between qubits and the environment. 
In addition, 
for most qubit systems, once the qubit array
is made, $\Omega_i$ and $J_{ij}$ are fixed, and only 
$\epsilon_i$ is controllable via the gate voltage bias 
(we call these ``simple-design qubits").   
%
Practically, such a simple design is preferable for solid-state qubits 
so as to simplify fabrication and enhance scalability. 
Thus, our goal here is to generate cluster states for simple-design
qubits.  Hereafter, without loss of generality, and for convenience, 
we focus on qubits of coupled
QDs.  $|0\rangle$ and $|1\rangle$ refer to the two single-qubit
states in which the excess charge is localized in the upper and
lower dot, respectively (see Ref.~\cite{tana0}).

Typically, cluster states are generated by an Ising-like
Hamiltonian $H_{\rm cs}\!=\!(g/4)\sum_{i<j}
(1\!-\!\sigma_{iz})(1\!-\!\sigma_{jz})$, where $i$ and $j$ are 
nearest-neighbor sites, starting from an initial state 
$|\Psi_0\rangle$.  Preparing
a unitary evolution $U_{\rm cs}(t)=\exp (itH_{\rm cs})$
(we use $\hbar=1$) at $gt= (2n_{\rm I}+1)\pi$, 
where $n_{\rm I}$ is an integer,  
is the first step for one-way quantum computing.
%
%
Since simple-design charge qubits have Ising-like interactions, 
all we need to do to get 
$H_{\rm cs}$ out of Eq.~(\ref{Hcq}) is to turn off the 
effect of the $\sigma_x$ terms in (\ref{Hcq}).
This can be achieved in the high-bias regime where
$\epsilon_i \gg \Omega_i > J_{ij}$, by
applying a canonical transformation to Eq.~(\ref{Hcq}) \cite{canonical}:
%
\begin{equation}
H^{(\rm eff)}_{\rm cq}\approx H_{\rm cq}+[S,H_{\rm cq}]
= \sum_i E_{i}\sigma_{iz}\!+\!
\sum_{i<j}J_{ij}\sigma_{iz}\sigma_{jz}+H_{\rm uw}
\label{H_charge}
\end{equation}
where $S=-i \sum_i [\Omega_i/(2\epsilon_i)] \sigma_{iy}$, with
$\Omega/(2\epsilon_i) \ll 1$.
Also, $E_i\!=\! \epsilon_i\!+\!\Omega_i^2/\epsilon_i$, 
neglecting terms of higher-order than $(\Omega_i/\epsilon_i)^2$.
$H_{\rm uw}$ is an unwanted interaction term given by
$H_{\rm uw}=-\sum_{i=1}^{N}\sum_{{\vec \gamma}_d} (\Omega_i/\epsilon_i)
J_{i,i+{\vec \gamma}_d} \, \sigma_{ix}\,\sigma_{i+{\vec \gamma}_d,z}$
for a $d$-dimensional qubit array.
Here ${\vec \gamma}_1=\pm 1$, ${\vec \gamma}_2=\{(\pm1,0),(0,\pm1)\} $
and ${\vec \gamma}_3=\{(\pm1,0,0),(0,\pm1,0),(0,0,\pm1)\}$ are
nearest neighbor indices for one-, two- and three-dimensional qubit arrays, 
respectively.
%
%
As long as $H_{\rm uw}$ is sufficiently small and can be neglected,
we can periodically generate cluster states 
in the tilted frame $|\tilde{\Psi}(t)\rangle=e^{-S}|\Psi(t)\rangle$
after a time $t_{\rm cs}$, 
if both $Jt_{\rm cs}=\pi/4+2n_{\rm J}\pi$ and
$E_i\,t_{\rm cs}=-(\pi/4)\bar{n}_i+2n_{\rm E}\pi$
are satisfied ($\bar{n}_i$
is the number of qubits connected to the $i$-th qubit;
$n_{\rm J}$ $(\ge 0)$ and $n_{\rm E}$ are arbitrary integers, 
and $J_{ij}$ should be uniform: $J_{ij}=J$).
These two equalities lead to the relation 
$J(8n_{\rm E}-\bar{n}_i)
/(8n_{\rm J}+1)=E_i\equiv E_{i,t_{\rm cs}}$.  Thus,
to generate a cluster state,
gate bias voltage $\epsilon_i$ for the $i$-th qubit needs to be set at
\begin{equation}
\epsilon_i = \epsilon_i^{\rm cs}=\frac{E_{i,t_{\rm cs}}}{2} 
\!+\!\sqrt{\left( \frac{E_{i,t_{\rm cs}}}{2} \right)^2
- \Omega_i^2}
\label{eqn:bias}
\end{equation}
during $t_{\rm cs}= \pi(8n_{\rm J}+1)/(4J)$. 
To ensure a solution for $\epsilon_i^{\rm cs}$ exists, 
we require $n_{\rm E}-(2\Omega_i/J)n_{\rm J} > (2\Omega_i/J+\bar{n}_i)/8$.
Hereafter we choose to treat the case of $n_{\rm J}=0$, which 
corresponds to the shortest possible time to generate cluster states. 
To further ensure the high-bias regime, where $\epsilon_i \gg \Omega_i$, 
we require $8n_{\rm E} - \bar{n}_i \gg 2\Omega_i/J$. 

An initial product state $|\tilde{\Psi}(0)\rangle =|\Psi_0\rangle$
has to be prepared
for qubits other than the input qubits \cite{Briegel2}.
$|\Psi(0)\rangle =\Pi_i(\cos [(\Omega_i/(2\epsilon_i)+\pi/4]\,|0\rangle
+\sin [(\Omega_i/(2\epsilon_i)+\pi/4]\,|1\rangle) $ is the corresponding
state in the original $\{|0\rangle, |1\rangle\}$ basis, which
can also be adjusted by the gate bias on each qubit. 

For example, to obtain the two-qubit cluster state $|\Psi\rangle_{C_2}=$ 
$(|0\rangle_1|+\rangle_2 \!+\! |1\rangle_1 |-\rangle_2)/\sqrt{2}$, 
where $\bar{n}_1\!=\!\bar{n}_2\!=\!1$, 
we can use Eq.~(\ref{eqn:bias}) and choose
a gate bias ($\epsilon_1^{\rm cs}\!=\!\epsilon_2^{\rm cs}$) from
$\{ 6.37J, 14.7J,...\}$
($n_{\rm E} \ge 1$) for $\Omega=2J$, and
$\{13.8J, 22.3J,...\}$
($n_{\rm E} \ge 2$) for $\Omega=4J$, etc.


Our approach is valid as long as the unwanted term $H_{\rm uw}$
can be neglected.  We can estimate a lifetime, beyond which we
lose the cluster state due to the presence of $H_{\rm uw}$, 
by calculating the {\it fidelity} defined by
$F(t)= \langle \Psi_0 |
e^{iH_{\rm cs}t}e^{-i(H_{\rm cs}+H_{\rm uw})t}
|\Psi_0\rangle
\approx 1- i t  \langle \Psi_0 |H_{\rm uw}|\Psi_0\rangle
\!+\!(1/2)\left(it \right)^2
\langle \Psi_0 | [H_{\rm uw},H_{\rm cs}]\!+\!H_{\rm uw}^2 |\Psi_0\rangle
$.
For a $d$-dimensional $N$-qubit array,
\begin{equation}
F(t)\approx
1-\left(\frac{\Omega}{\epsilon}Jt\right)^2 4d N
\end{equation}
Thus, the lifetime of the cluster state is limited by
$t<t_{\rm uw}\equiv(2J(\Omega/\epsilon)\sqrt{dN})^{-1}$. 
Furthermore, the constraint $t_{\rm cs} < t_{\rm uw}$ imposes 
a limit on the number of clustered qubits:
$N_{\rm max}<(2\epsilon/(\pi\Omega))^2/d$.
For example, consider a one-dimensional qubit-chain 
with $\Omega_i=4J$, and $\bar{n}_i=2$.
For $n_{\rm E}=2$ ($\epsilon_i^{\rm cs}/J\approx 12.7$), 
$N_{\rm max}=4$.  For $n_{\rm E}=6$ 
($\epsilon_i^{\rm cs}/J \approx 45.6$),  
$N_{\rm max}=52$.  Indeed, we can choose an infinite number of bias
conditions for each set of fixed $\Omega_i$ and $J$.  These are closely
related to the possible number $N_{\rm max}$ of clustered qubits and 
to the scalability of the system. 
%
%
Various kinds of errors, as discussed in
Ref.~\cite{Nielsen}, should be taken into account for more detailed estimates.

As shown in Ref.~\cite{tana0}, $J_{ij}$ and $\Omega_i$ are
determined by the distances between QDs. $J_{ij}$ is
directly determined by the capacitance network of QDs and
basically controlled as a linear function of the distances
between QDs.  $\Omega_i$ depends exponentially on the distances
between two QDs in a qubit.  With current technology it is quite difficult
to fabricate an array of QDs with very uniform $\Omega_i$.  
For superconducting charge qubits the situation is similar.
However, notice that our approach does not require $\Omega_i$ to be extremely
uniform since we can adjust $\epsilon_i$ according to $\Omega_i$ 
in order to obtain an appropriate $E_i$ in
Eq.~(\ref{H_charge}). 
%
%
In addition, as noted above, $n_{\rm E}$ can be
selected arbitrarily, which adds flexibility to our scheme.
%
%
Thus, the one-touch cluster state generation method we discuss here 
should work with any charge qubit architecture.  In short, although in general charge 
qubits have shorter decoherence times compared with spin qubits, 
our simpler and faster generation method could make them competitive with 
spin qubits in the context of cluster states, since several steps are 
required to generate cluster states for spin qubits \cite{Loss,Levy}.

{\it Measurement scheme in charge qubits.}--- In one-way quantum computing, 
calculations are carried out by
a series of local measurements in the $\sigma_x$ and $\sigma_y$ eigenbasis.
For most charge qubits, however, the
measurement is carried out in the
$\sigma_z$ eigenbasis $\{|0\rangle,|1\rangle\}$
by simply applying a large gate bias and using field-effect
detectors such as quantum point contacts (QPCs) or
single-electron transistors.
Thus, for charge qubits the $\sigma_x$ and $\sigma_y$ basis measurements
should be converted into $\sigma_z$ measurements after
rotating the frame via $\pm(\pi/2)_y$ and $\pm(\pi/2)_x$ pulses.
%
%
These pulses can be generated by applying
AC gate biases such as
$\epsilon_i (t)=\epsilon_{0i} \cos (\omega_{ic} t+\phi_i)$.
In a rotating coordinate
$U_{\rm rw} (t) =\exp (-i\sum_i \omega_{ic} t \sigma_{ix} )$,
the wave function is given by
$|\tilde{\Psi}(t) \rangle=U_{\rm rw}^\dagger |\Psi(t) \rangle$,
and the Hamiltonian on resonance ($\omega_{ic}=\Omega_i$) is given by:
\begin{equation}
H_{\rm rw}
\approx \sum_i \frac{\epsilon_{i0}}{2}
(\sigma_{iz} \,\cos \phi_i   -\sigma_{iy} \,\sin \phi_i)
+H_{yz}\;,
\label{eqn:rw}
\end{equation}
where $H_{yz}\!=\!\sum_{ij}(J_{ij}/2) (\sigma_{iy}\,
\sigma_{jy}+\sigma_{iz}\, \sigma_{jz})$ is an
unwanted term here.  In order to realize $ |\Psi(t_{\rm m})\rangle$
$=e^{\pm i\frac{\pi}{4}\sum_i \sigma_{iy}} |\Psi(0)\rangle $ at
the measurement time $t\!=\!t_{\rm m}$, we should have
$
e^{iH_{\rm rw}t_{\rm m}} \propto
e^{i\sum_i \Omega_i t_{\rm m} \sigma_{ix}}
e^{\mp i\frac{\pi}{4}\sum_i \sigma_{iy}}
$. Thus, we obtain $t_{\rm m}=(\pi/\Omega)l_1$,
the voltage amplitude $\epsilon_{i0}=(\pi/2t_{\rm m})(1+4l_2)$, and
the phase $\phi_i=\pi/2+l_3\pi$ ($l_1$, 
$l_2$ and $l_3$ are arbitrary integers ($l_1\neq0$))
for the $\sigma_x$ measurement.
If we take $t_{\rm m}=(\pi/(4\Omega))(1+4l_1)$ and 
$\epsilon_{i0}=2\pi l_2/t_{\rm m}$
($l_2\neq0$),
we realize the $\sigma_y$ measurement. 
By applying a gate bias after the time $t_{\rm m}$,  $\sigma_x$ and $\sigma_y$
measurements are processed.
This scheme works well for the region where $\epsilon_{i0}/J \gg 1$.

%
%
%
For the experiment in
Ref.~\cite{Yamamoto}, $\Omega=20.5~\mu$eV and $J=95~\mu$eV, 
and thus 
$t_{\rm cs}=\pi/(4J)
=34.2$ psec, 
$\epsilon_{\rm cs} \sim 664~\mu$eV from Eq.~(\ref{eqn:bias}) at
$n_{\rm E}=1$ and $t_{\rm uw}=(\pi/J)(\epsilon_{\rm cs}/[2\sqrt{2}\Omega])
%
%
=1.57$ nsec for a dephasing time of the order of $T_2\sim 5$ nsec.
%
%
Thus, our approach can be applied, if smaller $J$ ($<\Omega$) is prepared, 
for instance, by increasing the distance between qubits.  
%

{\it Cluster states in flux qubits.}--- For flux qubits, $\Omega >
\epsilon$ in Eq.~(\ref{Hcq}), so that here we cannot directly use the 
above-mentioned one-touch approach for charge qubits.
Here we show a method to generate cluster states
for flux qubits by applying an oscillating magnetic field.  Consider
two inductance-coupled flux qubits working at the optimal
bias~\cite{Rigetti} with Hamiltonian:
\begin{eqnarray}
H_{\rm fq}&=& \epsilon_1 \;\sigma_{1z}
+\epsilon_2 \;\sigma_{2z}
+\Omega_1^R \cos(\omega_1^{\rm rf}t+\phi_1)\,\sigma_{1x} 
\nonumber \\
&+&\Omega_2^R \cos(\omega_2^{\rm rf}t+\phi_2)\,\sigma_{2x}
+J_{xx}\;\sigma_{1x}\;\sigma_{2x}\ ,
\label{eqn:flux}
\end{eqnarray}
where $\Omega_i^R$ and $\omega_i^{\rm rf}$ are the half amplitude
and the frequency respectively of the applied classical field.
At the optimal bias point, the system is immune, up to first order, 
to variations on the control parameters and is thus robust against
decoherence.  This Hamiltonian is a good starting point
for generating cluster states.
In the rotating wave approximation
for two identical qubits ($\Omega_1=\Omega_2$), 
we have $\tilde{H}_{\rm fq}=H_{0}+H_{xy}$, with
$H_{0}=\sum_{i=1}^2 (\Omega^R_i/2)
(\sigma_{ix}\cos \phi_i+\sigma_{iy} \sin \phi_i)$
and $H_{xy}=J_{xx}(\sigma_{1x}\sigma_{2x} + 
\sigma_{1y}\sigma_{2y})$.  The operator to generate cluster states,
$U_{\rm cs}$, is produced by switching on and off the resonant
field of $\Omega_i^R$ and controlling the phase $\phi_i$
similarly to the conventional {\it conditional phase gate} operation.
For example, if we define $R_{i\alpha}(\theta)\equiv \exp(i\theta 
\sigma_{i\alpha})$, ($\alpha=x,y$), and
$U_{xy}(\theta)\equiv \exp(i\theta (\sigma_{1x}\sigma_{2x}
+\sigma_{1y}\sigma_{2y}))$, we have:
\begin{eqnarray}
\tilde{U}_{\rm cs}\!&\!\!=\!&\!\!R_{1x}\!\left(\theta_1\!\right)
\!R_{2x}\!\left(\theta_2\!\right)\!
U_{xy}\!\left(\theta_3\!\right)\!R_{1x}\!\left(\theta_4\!\right)\!
U_{xy}\!\left(\theta_5\!\right)\!R_{1x}\!\left(\theta_6\!\right)
\nonumber \\
&=&\exp \left({-i\pi(\sigma_{1x}+\sigma_{2x}-\sigma_{1x}\sigma_{2x})/4}\right)
\label{U_flux}
\end{eqnarray}
with
$\theta_1=\theta_2=-\pi/4$,
$\theta_3=\theta_5=\pi/8$,
$\theta_4=\pi/2$ and $\theta_6=-\pi/2$.
After rotating $\tilde{U}_{\rm cs}$ around the $y$-axis, 
we recover the original cluster state generator $U_{\rm cs}$.
In the case of many qubits, cluster states for the entire system 
are generated by applying Eq.~(\ref{U_flux}) to all the neighboring 
qubit pairs.  Note
that Eq.~(\ref{eqn:flux}) also describes the
rotating wave approximation to the charge qubits in Eq.~(\ref{eqn:rw}).
Thus, this method of generating cluster states is also 
applicable to charge qubits, and not only to flux qubits.

The time required for the creation of a cluster state in flux qubits 
is $ T_{\rm flux}=5\pi/(2\Omega_1^R)+\pi/(4\Omega_2^R) +\pi/(4J_{xx}) $.
Taking $\Omega_1^R\sim\Omega_2^R\sim J_{xx}\sim$ 0.5 GHz,
we obtain $T_{\rm flux}\sim 18$ nsec ($T_2\sim$ 200
nsec~\cite{Liu}).  The effect of imperfect pulses can be
estimated by substituting $\theta_j \rightarrow
\theta_j+\delta_j$ in Eq.~(\ref{U_flux}).  If we take the
deviation from a perfect pulse as $\delta_1=\delta_2$,
$\delta_3=\delta_5$, $\delta_4=\delta_6$, for the initial state
$|\Psi_0\rangle$, the {\it fidelity}
is given by
\begin{equation}
F(t)\approx
1-\delta_1^2/2-2\,\delta_3^2 +\sin^2 [\pi/8] 
\,(\delta_4^2/2)
+i\delta_1\delta_4 \,e^{i\pi/8}
\end{equation}
Thus, the fidelity $F(t)$ remains close to one, up to second order in the 
pulse shape error, even when the pulse shape has defects.

\begin{figure}
\begin{center}
\includegraphics[width=8.5cm]{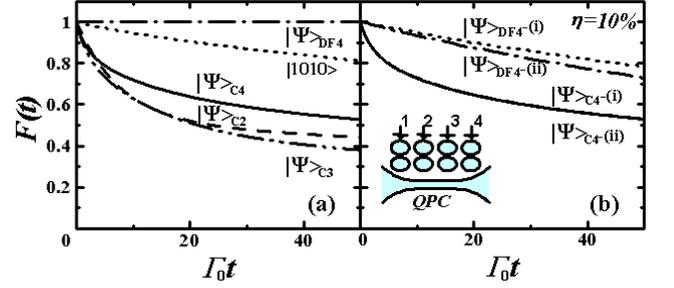}
\end{center}
\caption{Time-dependent fidelities $F(t)$
of cluster states $|\Psi\rangle_{\rm C2}$,
$|\Psi\rangle_{\rm C3}$, $|\Psi\rangle_{\rm C4}$
, a product state $|1010\rangle$ 
and four-qubit decoherence-free (DF) states $|\Psi\rangle_{\rm DF4}
=(|1100\rangle
-|1001\rangle
-|0110\rangle
+|0011\rangle)/2$,
for $\Gamma_0=J$ and $\Delta \Gamma=0.6J$. $\Omega=4J$ and $n_{\rm E}=2$
in Eq.~(\ref{eqn:bias}), thus $\epsilon_2=\epsilon_3\sim 12.7J$,
$\epsilon_1=\epsilon_4\sim 13.8J$. (a) Comparison of the cluster
states, DF state and $|1010\rangle$. (b) The case when non-uniformity in the 
qubit parameters is introduced as
$\Omega_i\!=4J \!(1\!-\!\eta_i)$, $\epsilon_i\!=\epsilon+\!\eta_i
J$ and $\Gamma_i^{(\pm)}\!=\!(1\!-\!\eta_i)\Gamma^{(\pm)}$, with
$i$ indicating the $i$-th qubit. 
Here $\eta_i=0$ for all qubits besides $\eta_3=0.1$ (i) and $\eta_4=0.1$ (ii).
The fidelities of
$|\Psi\rangle_{\rm C4}$ for (i) and (ii) mostly overlap.
(inset) Four qubits that use double dot charged states are
capacitively coupled to a QPC detector. We consider a similar
detection setup for two and three qubits.
These calculations are carried out by the $H_{\rm cq}$,
that is, they include  $H_{\rm uw}$ and higher order terms.
} \label{FID}
\end{figure}

{\it Effect of non-uniformity in cluster states.}--- Cluster states are
highly entangled states involving all qubits, and can be decohered by 
%
%
various kinds of {\it local} fluctuations.  Here we
investigate the effect of non-uniform qubit parameters on cluster states 
in semiconductor QDs, from Eq.~(\ref{Hcq}), using a 
measurement setup, which produces decoherence in the double dot
qubits through back-action. 
%
%
We analyze a capacitively coupled detector (such as a QPC), whose 
shot-noise constitutes a random charge fluctuation on the qubits.

We use a density matrix (DM) to describe up to four qubits 
(inset of Fig.\ref{FID}) \cite{tana}.
The DM equations for the qubits and the QPC detector are derived in
Ref.~\cite{tana}:
\begin{eqnarray}
\frac{d \rho_{z_1,z_2}}{dt}&\!=\!&i[K_{z_2}\!-\!K_{z_1}\!]
\rho_{z_1,z_2}
\!-\! i\sum_{j=1}^{N} \Omega_j (\rho_{g_j(z_1),z_2}\!-\!\rho_{z_1,g_j(z_2)})
\nonumber \\
&\!-\!&\left[ \Gamma_{z_1}^{1/2}\!-\!\Gamma_{z_2}^{1/2}\right]^2
\rho_{z_1,z_2} 
\label{eqn:dm}
\end{eqnarray}
\normalsize
where $z_1,z_2=(1111), (1110), ...,
(0000)$ for four qubits (256 equations) and
$z_1,z_2=(11), (10), (01), (00)$
for two qubits (16 equations).
%
%
$K_{z_1}$ is the energy of the $z_1$ state and depends on $\epsilon_i$ 
and $J_{ij}$ in Eq.~(\ref{Hcq}).  For example, for two qubits, 
$K_{(11)} \!=\! \epsilon_1\!+\!\epsilon_2\!+\!J_{12}$
while $K_{(10)} \!=\! \epsilon_1\!-\!\epsilon_2\!-\!J_{12}$.
%
%
The $g_j(z_i)$s are introduced for notational
convenience and are determined by the relative positions between qubit
states.
We assume that the tunneling rate $\Gamma$ of the 
QPC detector in the presence of $N$ qubits satisfies 
$\Gamma^{-1}\!=\!\sum_{i} \Gamma_{i}^{-1}$,  
where the tunneling rate $\Gamma_{i}$ is determined by
the state $\sigma_{iz}\!=\pm \,1$ of the $i$-th qubit.
The strength of the measurement can be parametrized by 
$\Delta \Gamma_{i}$
as $\Gamma_{i}^{(\pm)}\!=\!\Gamma_{i0}\,\pm\Delta \Gamma_{i}$, 
where $\Gamma_{i0}$ is the tunneling rate of the QPC in the 
absence of the qubits. 
The time-dependent {\it fidelity}
$F(t)\!\equiv\! {\rm Tr}[\hat{\rho}(0) \hat{\rho}(t)]$
can be calculated by tracing over the elements 
of the reduced DM obtained from Eq.~(\ref{eqn:dm}).
$F(t)$ can be expanded in time as $F(t)\!=\!1\!-\!\sum_{n\!=\!1}
(1/n!)(t/\tau^{(n)})^n$, where the lifetime is
$1/\tau^{(n)} =\{-{\rm Tr}[\hat{\rho}(0)
d^n\hat{\rho}(0)/dt^n]\}^{1/n}$.

From Eq.~(\ref{eqn:dm}), we obtain the first-order lifetime for
two-, three- and four-qubit cluster states $|\Psi\rangle_{C_2}$,
$|\Psi\rangle_{C_3}=(|+\rangle_1 |0\rangle_2 |+\rangle_3 \!
+\!|-\rangle_1 |1\rangle_2|-\rangle_3)/\sqrt{2}$, and
$|\Psi\rangle_{C_4}=(|+\rangle_1 |0\rangle_2
|+\rangle_3|0\rangle_4 \!+\!|+\rangle_1 |0\rangle_2
|-\rangle_3|1\rangle_4 \!+\!|-\rangle_1 |1\rangle_2
|-\rangle_3|0\rangle_4 \!+\!|-\rangle_1 |1\rangle_2
|+\rangle_3|1\rangle_4) /2$, respectively, as follows:
\begin{eqnarray}
1/\tau^{(1)}_{C2}&=&\sum_{z_1,z_2=(11),..,(00)}
{\it \Gamma}_{\rm d}(z_1,z_2)/8  \\
1/\tau^{(1)}_{C3}&=&\sum_{z_1,z_2=(111)
,..,(000)}{\it \Gamma}_{\rm d}(z_1,z_2)/32, \\
1/\tau^{(1)}_{C4}&=&\sum_{z_1,z_2=(1111)
,..,(0000)}
{\it \Gamma}_{\rm d}(z_1,z_2)/128,
\end{eqnarray}
where the dephasing rate is defined as
${\it \Gamma}_{\rm d}(z_1,z_2)$
$\equiv\![\Gamma_{z_1}^{1/2}\!-\!\Gamma_{z_2}^{1/2}]^2$.
Note that the lifetime of the cluster states is an average
over all the dephasing rates between different product states.
This is in contrast with other entangled states.
For example, the lifetime of two-qubit Bell states 
$|c\rangle =(|10 \rangle+|01 \rangle)/\sqrt{2}$
and $|d\rangle =(| 10 \rangle-| 01 \rangle)
/\sqrt{2}$ takes the form
$1/\tau^{(1)}_c\!=\!1/\tau^{(1)}_d
\!=\!(1/2){\it \Gamma}_{\rm d}(10,01)$.
It is well known that the singlet state $|d\rangle$ is the most robust 
two-qubit state \cite{Zanardi} when there is a symmetry between 
qubits.  However, solid-state qubits generally 
decohere due to various kinds of local causes, 
which often break the symmetry of the qubit state. 
Our results in Eq.~(10-12) indicate that cluster states might be
robust against non-uniformity or local
defects because of the averages.

In Fig.~\ref{FID} we compare the fidelity of cluster states with a product 
state and a four-qubit decoherence-free (DF) state~\cite{Zanardi}.
Figure \ref{FID}(a) shows the time-dependent fidelities of 
two-qubit and four-qubit cluster states
and the product state $|1010\rangle$, when $\Gamma_0=J$ and $\Delta \Gamma=0.6J$.
Our results show that the strongly entangled cluster states
are more fragile than a product state such as $|1010\rangle$. 
We can also see that the robustness of the cluster state
depends on the number of qubits in the cluster state.
Figure~\ref{FID}(b) shows the time-dependent fidelities of both 
cluster states and DF states,  
for non-uniform qubits.  Here the parameters $\Omega_i$, $\epsilon_i$ and 
$\Gamma_i$ for the third or fourth qubit deviate from those of other qubits
by 10\%.  Note that the fidelities of $|\Psi\rangle_{\rm C4}$
show almost the same behavior irrespective of the distribution
of the non-uniformity.  Furthermore, a comparison between 
Fig.~\ref{FID}(a) and ~\ref{FID}(b) shows that the non-uniformity has 
almost no effect on the fidelity of the $|\Psi\rangle_{\rm C4}$ cluster
state.  These results vividly illustrate 
our analytical analysis of the lifetime (that it is an average over all
the product states).  In contrast, in
Ref.~\cite{tana} we showed that the robustness of the DF states
strongly depends on the non-uniformity.  Thus, even though 
cluster states are generally more fragile  
than DF states, they are more robust against non-uniformities 
among qubits than DF states.

In conclusion, we describe how to efficiently generate cluster states 
in solid-state qubits.  By manipulating the gate bias voltage, 
we explicitly show how to generate `one-touch' entanglement
via cluster states in charge qubits.
We also investigate the robustness of cluster states, and
find that one-way quantum computing could be viable 
for solid-state qubits.


FN and XH are supported in part by NSA, LPS, ARO, and NSF.



\begin{thebibliography}{99}
\bibitem{canonical}
We take the first order of the expansion 
$e^{-S}H_{\rm cq}e^{S}
=H_{\rm cq}+[S,H_{\rm cq}]+[S,[S,H_{\rm cq}]]/(2!)...$
\end{thebibliography}
\end{document}